\documentclass[twocolumn]{bmcart}

\usepackage[figure,noline]{algorithm2e}
\usepackage{anyfontsize}
\usepackage{amsthm,amsmath}
\usepackage{colortbl}
\usepackage{graphicx}
\usepackage{hyperref}
\usepackage[utf8]{inputenc} 
\usepackage{lmodern}
\usepackage{makecell}
\usepackage{multirow}
\usepackage{siunitx}
\usepackage{threeparttable}
\usepackage{wasysym}

\graphicspath{{./}{figures/}}	

\startlocaldefs%
\DeclareMathOperator*{\argmax}{arg\,max}%
\DeclareMathOperator*{\coords}{coords}%
\endlocaldefs%

\begin{document}

\begin{frontmatter}

\begin{fmbox}
  \dochead{Research}


  \title{A Scalable Deep Neural Network Architecture for Multi-Building and
    Multi-Floor Indoor Localization Based on Wi-Fi Fingerprinting}


  \author[ addressref={aff1,aff2}, 
  corref={aff1}, 
  email={Kyeongsoo.Kim@xjtlu.edu.cn} 
  ]{\inits{KS}\fnm{Kyeong Soo} \snm{Kim}}

  \author[ addressref={aff1,aff2}, 
  email={Sanghyuk.Lee@xjtlu.edu.cn} 
  ]{\inits{S}\fnm{Sanghyuk} \snm{Lee}}

  \author[ addressref={aff1}, 
  email={Kaizhu.Huang@xjtlu.edu.cn} 
  ]{\inits{K}\fnm{Kaizhu} \snm{Huang}}


  \address[id=aff1]{
    \orgname{Department of Electrical and Electronic Engineering,
      XJTLU}, 
    \street{111 Ren'ai Road}, \postcode{215123} 
    \city{Suzhou}, 
    \cny{China} 
  }

  \address[id=aff2]{
    \orgname{ Centre for Smart Grid and Information Convergence,
      XJTLU}, 
    \street{111 Ren'ai Road}, \postcode{215123} 
    \city{Suzhou}, 
    \cny{China} 
  }


  \begin{artnotes}
    An earlier version of this paper was presented in part at ICUMT/FOAN 2017,
    Munich, Germany, November 2017.
  \end{artnotes}

\end{fmbox}


\begin{abstractbox}

  \begin{abstract}
    One of the key technologies for future large-scale location-aware services
    covering a complex of multi-story buildings --- e.g., a big shopping mall
    and a university campus --- is a \textit{scalable indoor localization
      technique}. In this paper, we report the current status of our
    investigation on the use of deep neural networks (DNNs) for scalable
    building/floor classification and floor-level position estimation based on
    Wi-Fi fingerprinting. Exploiting the hierarchical nature of the
    building/floor estimation and floor-level coordinates estimation of a
    location, we propose a new DNN architecture consisting of a stacked
    autoencoder for the reduction of feature space dimension and a feed-forward
    classifier for \textit{multi-label classification} of
    building/floor/location, on which the multi-building and multi-floor indoor
    localization system based on Wi-Fi fingerprinting is built. Experimental
    results for the performance of building/floor estimation and floor-level
    coordinates estimation of a given location demonstrate the feasibility of
    the proposed DNN-based indoor localization system, which can provide near
    state-of-the-art performance using a single DNN, for the implementation with
    lower complexity and energy consumption at mobile devices.
  \end{abstract}




\end{abstractbox}
%

\end{frontmatter}



\section{Introduction}
\label{sec:introduction}
\textit{Location fingerprinting} using received signal strengths (RSSs) from
wireless network infrastructure is one of the most popular and promising
technologies for localization in an indoor environment, where there is no
line-of-sight signal from the global positioning system (GPS) available
\cite{he16:_wi_fi}: For example, a vector of pairs of a service set identifier
(SSID) and an RSS for a Wi-Fi access point (AP) measured at a location can be
its \textit{location fingerprint}. A location of a user/device then can be
estimated by finding the closest match between its RSS measurement and the
fingerprints of known locations in a database \cite{bahl00:_radar}. Note that
the location fingerprinting technique does not require the installation of any
new infrastructure or the modification of existing devices, but it is just based
on the existing wireless infrastructure, which is its major advantage over
alternative techniques.

When the indoor localization is to cover a large \textit{building complex} ---
e.g., a big shopping mall or a university campus --- where there are lots of
multi-story buildings under the same management, the scalability of
fingerprinting techniques becomes an important issue. The current
state-of-the-art Wi-Fi fingerprinting techniques assume a hierarchical approach
to the indoor localization, where the building, floor, and position (e.g., a
label or coordinates) of a location are estimated in a hierarchical and
sequential way using a different algorithm tailored for each task. In
\cite{moreira15:_wi_fi}, for instance, building estimation is done as follows:
Given the AP with the strongest RSS in a measured fingerprint, we first build a
subset of fingerprints where the same AP has the strongest RSS; then, we count
the number of fingerprints associated to each building and set the estimated
building to be the most frequent one from the counting. Similar procedures are
also proposed to estimate a floor inside the building. For the estimation of the
coordinates of the location, we first build a subset of fingerprints belonging
to the building and the floor estimated from the previous procedures. Then, take
multiple fingerprints from the subset most similar to the measured one, and
compute the centroid of the coordinates of the selected fingerprints as the
estimated coordinates of the given location. According to the results in
\cite{moreira15:_wi_fi}, the best building and floor hit rates achieved for the
UJIIndoorLoc dataset \cite{torres-sospedra14:_ujiin} are 100\% and 94\%,
respectively, and the mean error in coordinates estimation is
\SI{6.20}{\meter}.\footnote{The mean error takes into account the building and
  floor estimation penalties; refer to \cite[Eq.~(2)]{moreira15:_wi_fi} for
  details.}

One of the major challenges in Wi-Fi fingerprinting is how to deal with the
random fluctuation of a signal, the noise from multi-path effects, and the
device \& position dependency in RSS measurements. Unlike traditional solutions
relying on complex filtering and time-consuming parameter tuning specific to
given conditions, the popular deep neural networks (DNNs) can provide attractive
solutions to Wi-Fi fingerprinting due to their less parameter tuning and
adaptability to a wider range of conditions with standard architectures and
training algorithms \cite{zhang16:_deep,felix16,nowicki17:_low_wifi}: In
\cite{zhang16:_deep}, a four-layer DNN generates a coarse positioning estimate,
which, in turn, is refined to produce a final position estimate by a hidden
Markov model (HMM)-based fine localizer. The performance of the proposed indoor
localization system is evaluated in both indoor and outdoor environments which
are divided into hundreds of square grids. In \cite{felix16}, the authors
investigate the application of deep belief networks (DBNs) with two different
types of Restricted Boltzmann Machines for indoor localization and evaluate the
performance of their approaches using data from simulation in heterogeneous
mobile radio networks using ray tracing techniques. In both cases, the authors
focus only on the localization in a single plane and do not consider the
hierarchical nature of multi-building and multi-floor indoor localization. In
\cite{nowicki17:_low_wifi}, on the other hand, a DNN consisting of a stacked
autoencoder (SAE) and a feed-forward multi-class classifier is used for
building/floor classification. This work, too, does not take into account the
hierarchical nature of building/floor classification, because the classification
is done over flattened, one-dimensional labels of combined building and floor
identifiers. Also, the floor-level location estimation is not considered at
all. In this regard, to the best of our knowledge, the work presented in this
paper is the first to apply DNNs for multi-building and multi-floor indoor
localization, exploiting its hierarchical nature in classification.

In this paper, we report the current status of our investigation on the use of
DNNs for scalable building/floor classification and floor-level location
estimation. We propose a new DNN architecture consisting of an SAE for the
reduction of feature space dimension and a feed-forward \textit{multi-label
  classifier} \cite{tsoumakas07:_multi,herrera16:_multil} for a multi-building
and multi-floor indoor localization system based on Wi-Fi fingerprinting and
evaluate its performance using the UJIIndoorLoc dataset
\cite{torres-sospedra14:_ujiin}.

The rest of the paper is organized as follows: In Sec.~\ref{sec:scal-dnn-arch},
we describe the problem of indoor localization in a large building complex with
its challenges resulting from the existence of multi-story buildings and propose
a scalable DNN architecture for the multi-building and multi-floor indoor
localization. Sec.~\ref{sec:experimental-results} provides and discusses
experimental results for the performance of the proposed scalable DNN-based
multi-building and multi-floor indoor localization
system. Sec.~\ref{sec:conclusions} concludes our work in this paper and suggests
areas of further research.

\section{A Scalable DNN Architecture for Multi-Building and Multi-Floor Indoor
  Localization}
\label{sec:scal-dnn-arch}
Location awareness is one of enabling technologies for future smart and green
cities; understanding where people spend their times and how they interact with
environments is critical to realizing the vision of smart and green cities
\cite{benevolo16:_smart}. One of the key technologies for future large-scale
location-aware services covering a complex of multi-story buildings --- e.g., a
big shopping mall and a university campus --- is a scalable indoor localization
technique. Regarding the scalability of the indoor localization, consider the
evolution of the Xi'an Jiaotong-Liverpool University (XJTLU) campus in Suzhou,
China, where the authors are currently working: As shown in
Figure~\ref{fig:xjtlu_campus}~(a), the XJTLU started with just one building in
2006. As of this writing, the XJTLU has two campuses, which are shown in
Figure~\ref{fig:xjtlu_campus}~(b), and the number of buildings over two campuses
has increased to around 20; this number is still increasing as more buildings
and sports facilities are being constructed. Considering all the floors within
each building and the locations on each floor, the total number of distinct
locations (e.g., offices, lecture rooms, and labs) is already on the order of
thousands. If we adopt a grid-based representation of the localization area as
in \cite{zhang16:_deep}, the total number of locations would be even
greater. The indoor localization system to cover such a large building complex,
therefore, must be scalable.

Figure~\ref{fig:hierarchical_vs_integrated} shows two alternative system
architectures for large-scale DNN-based multi-building and multi-floor indoor
localization. In the hierarchical architecture shown in
Figure~\ref{fig:hierarchical_vs_integrated}~(a), the task of
building/floor/location classification is separated into multiple sub-tasks
dedicated to the classification at each level of building, floor, and
location. This architecture directly corresponds to the state-of-the-art
hierarchical Wi-Fi fingerprinting methods (e.g., \cite{moreira15:_wi_fi}), where
DNNs replace traditional techniques for building, floor, and location
estimation. Compared to the methods based on traditional techniques, a major
disadvantage in this hierarchical DNN architecture is that the DNNs in the floor
and the location levels of the system need to be trained separately with
multiple sub-datasets derived from a common dataset (i.e., building-specific
datasets for DNNs for floor estimation and building-floor-specific datasets for
DNNs for location estimation), which poses significant challenges on the
management of location fingerprint databases as well as the training of possibly
a large number of DNNs. In this paper, therefore, we focus on the integrated
architecture shown in Figure~\ref{fig:hierarchical_vs_integrated}~(b) where a
single DNN handles the classification of building, floor, and location in an
integrated way with a common dataset.

Figure~\ref{fig:flat_dnn_classifier} shows a DNN architecture for the combined
estimation of building, floor, and location based on \textit{multi-class
  classification} with flattened labels, which is a straightforward extension of
the DNN system for building/floor classification proposed in
\cite{nowicki17:_low_wifi};
after training with RSSs as both input and output data as shown in
Figure~\ref{fig:flat_dnn_classifier}~(a), only the gray-colored nodes are used as
an encoder for feature space dimension reduction as shown in
Figure~\ref{fig:flat_dnn_classifier}~(b). This DNN architecture based on
multi-class classification with flattened labels, however, has the scalability
issue that the number of output nodes is equal to the number of locations over
the building complex: In case of the UJIIndoorLoc dataset, the number of
distinct locations (i.e., also called \textit{reference points} in
\cite{torres-sospedra14:_ujiin}) over 3 buildings with 4 or 5 floors is 933.  It
also does not take into account the hierarchical nature of the
building/floor/location classification problem due to its calculating the loss
and the accuracy over \textit{flattened building/floor/location
  labels}\footnote{For example, we can form a flattened label
  ``B$_{i}$-F$_{i,j}$-L$_{i,j,k}$'' by combining a building, a floor, and a
  location label, where B$_{i}$, F$_{i,j}$, and L$_{i,j,k}$ denote the $i$th
  building, $j$th floor of the building, and $k$th location on the floor,
  respectively.}; the misclassification of building, floor, or location has
equal loss during the training phase. To reflect the hierarchical nature of the
building/floor/location classification in a DNN classifier, one can use a
hierarchical loss function --- e.g., a loss function with different weights for
building, floor, and location --- with the existing multi-class classifier and
flattened labels. Because the hierarchical loss function for flattened labels is
quite complicated and does not provide a closed-form gradient function, however,
training the DNN with the usual backpropagation procedure could be challenging.

To address the scalability issue of the DNN classifier based on multi-class
classification and take into account the hierarchical nature of the
building/floor/location classification, we propose a scalable DNN architecture
based on \textit{multi-label classification}\footnote{In multi-class
  classification (also called \textit{single-label classification}), an instance
  is associated with only a single label from a set of disjoint labels; in
  multi-label classification, on the other hand, an instance can be associated
  with multiple labels \cite{tsoumakas07:_multi}.} shown in
Figure~\ref{fig:scalable_dnn_classifier}. The building/floor/location
classification with the proposed architecture is done as follows: First,
building, floor, and location identifiers are mapped to sequential numbers, the
latter two of which are meaningful only in combination with higher-level
numbers; those numbers are one-hot encoded \textit{independently} and combined
together into a vector as a categorical variable for multi-label classification
as illustrated in Table.~\ref{tab:label_formation}. Then, the output vector from
the multi-label classifier is split into a building, a floor, and a location
vector by indexes as shown in Figure~\ref{fig:scalable_dnn_classifier}. Finally,
we estimate the building and the floor of a location as the index of a maximum
value of the corresponding vector through the $\argmax$ function. For the
estimation of the location coordinates, we select $\kappa$ largest elements from
the location vector (i.e.,
$\boldsymbol{L}{=}\big(\text{L}_{1},\ldots,\text{L}_{\max(\ldots,N_{L}(i.j))}\big)$
in Figure~\ref{fig:scalable_dnn_classifier}), filter out the elements whose
values are less than $\sigma{\times}\max(\boldsymbol{L})$ ($\sigma{\in}[0, 1]$),
and calculate the estimated coordinates of the location as either the normal or
weighted (with the values of the elements as weights) centroid of the remaining
elements as described in detail in Figure~\ref{fig:coordinates_estimation}.

Note that there are two design parameters --- i.e., $\kappa$ and $\sigma$ --- in
the location coordinates estimation procedure, the rationale of which is
illustrated in Figure~\ref{fig:coordinates_examples}: If we use only $\kappa$ as
a design parameter as in \cite{moreira15:_wi_fi} and sets its value to 5 in
Figure~\ref{fig:coordinates_examples}~(a), we can include the reference points
quite close to the new location (i.e., those inside the dotted circle) in the
estimation procedure and can generate good estimation. In
Figure~\ref{fig:coordinates_examples}~(b), however, the same value of $\kappa$
could result in poor estimation because the reference points 4 and 5 have to be
considered during the estimation. With $\kappa{=}3$, on the other hand, we can
expect good estimation with Figure~\ref{fig:coordinates_examples}~(b) but not
with Figure~\ref{fig:coordinates_examples}~(a) this time. If we can use both
$\kappa$ and $\sigma$ as design parameters, however, we can include good
reference points by properly setting $\sigma$ for a threshold value. The actual
effects of these design parameters on the location coordinates estimation are
investigated in Sec.~\ref{sec:experimental-results}.

As for the scalability of the proposed DNN architecture, the number of output
nodes becomes much smaller than that of the DNN architecture based on
multi-class classification: The number of output nodes for multi-label
building/floor/location classification is given by
\begin{align}
  \label{eq:output_node_number_multi-label}
  N_{B} & + \max\Big(N_{F}(1),\ldots,N_{F}(N_{B})\Big) \\ \nonumber
        & +
          \max\Big(N_{L}(1,1),\ldots,N_{L}\big(N_{B},N_{F}(N_{B})\big)\Big) ,
\end{align}
where $N_{B}$, $N_{F}(i)$, and $N_{L}(j,k)$ are the number of buildings in the
complex, the number of floors in the $i$th building ($i=1,\ldots,N_{B}$), and
the number of locations on the $k$th floor of the $j$th building
($j=1,\ldots,N_{B}$; $k=1,\ldots,N_{F}(j)$), respectively. Note that for
multi-class building/floor/location classification, the number becomes
\begin{equation}
  \label{eq:output_node_number-multi-class}
  \sum_{i=1}^{N_{B}}\sum_{j=1}^{N_{F}(i)}N_{L}(i,j) .
\end{equation}
According to \eqref{eq:output_node_number_multi-label}, the number of output
nodes of the proposed DNN architecture for the publicly available UJIIndoorLoc
dataset at the University of California, Irvine (UCI), Machine Learning
Repository\footnote{\url{https://archive.ics.uci.edu/ml/datasets/ujiindoorloc}.}
is given by 118 (i.e., the sum of the number of buildings (3), the maximum of
the numbers of floors of the buildings (5), and the maximum of the numbers of
locations\footnote{In the UJIIndoorLoc dataset, the position of a location is
  uniquely determined by four identifiers, i.e., \textit{BuildingID},
  \textit{Floor}, \textit{SpaceID}, and \textit{RELATIVEPOSITION}. For
  convenience, we combine the SpaceID and the RELATIVEPOSITION into one and
  mention it as \textit{location} throughout the paper so that the three
  identifiers for building, floor, and location uniquely determine the position
  of a location.} on the floors (110)), which is smaller than the number of
output nodes of the DNN architecture based on multi-class classification (i.e.,
905\footnote{There are slight differences between the statistics of the
  UJIIndoorLoc dataset described in \cite{torres-sospedra14:_ujiin} and those of
  the publicly available dataset at the UCI Machine Learning Repository.}). Note
that the difference could be much larger if the UJIIndoorLoc dataset covers all
the buildings on the Jaume I University (UJI) campus where the data were
collected.

Also, due to the clear mapping between building, floor, and location identifiers
and its corresponding one-hot-encoded categorical variable for the DNN-based
multi-label classifier, it is easy to carry out different processing for parts
of DNN outputs specifically for building, floor, and location as illustrated in
Figure~\ref{fig:scalable_dnn_classifier}. Especially, the use of multiple
elements in estimating location coordinates is a huge advantage in terms of
computational complexity because trained DNNs can generate multi-dimensional
output values in parallel; in traditional approaches, on the other hand,
selecting nearest locations based on Euclidean distances are complex and time
consuming. This flexibility in handling DNN outputs also makes it easy to apply
different weights to the cost of building, floor, and location classification
error during the training phase.

\section{Experimental Results}
\label{sec:experimental-results}
We carried out experiments using the UJIIndoorLoc dataset
\cite{torres-sospedra14:_ujiin} to evaluate the performance of the proposed
DNN-based multi-building and multi-floor indoor localization
system.\footnote{Source code for the implemented DNN models based on Keras
  \cite{keras} and TensorFlow \cite{tensorflow} is available online:
  \url{http://kyeongsoo.github.io/research/projects/indoor_localization/index.html}.
} We focus on the effects of the number of largest elements from the output
location vector (i.e., $\kappa$) and the scaling factor for a threshold (i.e.,
$\sigma$) in the location coordinates estimation procedure described in
Sec.~\ref{sec:scal-dnn-arch}. Table~\ref{tab:scalable_localization_dnn_parameters}
summarizes DNN parameter values for the experiments, which are chosen
experimentally and used throughout the experiments.

As indicated in \cite{nowicki17:_low_wifi}, the publicly available UJIIndoorLoc
dataset includes training and validation data, but not testing data which were
provided only to the competitors at the Evaluating Ambient Assisted Living
(EvAAL) competition at the International Conference on Indoor Positioning and
Indoor Navigation (IPIN) 2015 \cite{moreira15:_wi_fi}. Also, unlike the training
data, the validation data do not include location information (i.e.,
\textit{SpaceID} and \textit{RELATIVEPOSITION} fields) because the measurements
were taken at arbitrary points as would happen in a real localization system. In
this regard, we split the training data into new training and validation data
with the ratio of 70:30 for DNN training and validation with
building/floor/location labels for both. During the evaluation phase, the output
from the trained DNN are post-processed as described in
Sec.~\ref{sec:scal-dnn-arch} and compared with the building, floor, and
coordinates of a given location. In this way, we can compare out results of
multi-building and multi-floor indoor localization with the baseline and the
best results from \cite{torres-sospedra14:_ujiin} and \cite{moreira15:_wi_fi}.

Table~\ref{tab:scalable_localization_performance} summarizes our experimental
results, which show the effects of the number of largest elements from the
output location vector ($\kappa$) and the scaling factor for a threshold
($\sigma$) on the performance of multi-building and multi-floor indoor
localization. We highlight in light gray the rows with success rate (i.e., the
rate of successful estimation of both building and floor) higher than 90 \% and
positioning errors (both normal and weighted centroid) lower than
\SI{10}{\meter}, respectively.

In general, $\sigma$ in the range of \numrange[range-phrase = --]{0.1}{0.3}
produces the best localization performance for $\kappa{\leq}8$; once $\kappa$
becomes larger than 8, however, higher values of $\sigma$ (i.e., 0.4 for
$\kappa{=}9$ and 0.5 for $\kappa{=}10$) generate better performance. Considering
the coordinates location estimation examples shown in
Figure~\ref{fig:coordinates_examples} with their explanations in
Sec.~\ref{sec:scal-dnn-arch} regarding the use of two design parameters, we can
explain these results as follows: With a larger value of $\kappa$ (i.e., 9 and
10), there could be a higher chance of including reference points relatively far
from the given location as shown in
Figure~\ref{fig:coordinates_examples}~(b). In such a case, a tighter threshold
(i.e., a larger value of $\sigma$\footnote{When $\sigma{=}0$, there is no
  filtering (i.e., including all $\kappa$ reference points); when $\sigma{=}1$,
  only the reference point with the largest value is considered during the
  location coordinates estimation.}) can filter out those reference points.

According to the results shown in
Table~\ref{tab:scalable_localization_performance}, collectively the best results
are achieved when $\kappa{=}8$ and $\sigma{=}0.2$, which are highlighted in
gray. These results from the proposed DNN-based multi-building and multi-floor
indoor localization system --- i.e., 99.82~\% for building hit rate, 91.27~\%
for floor hit rate, 91.18~\% for success rate and \SI{9.29}{\meter} for
positioning error --- are favorably comparable to the baseline results --- i.e.,
89.92 \% for success rate and \SI{7.9}{\meter} for positioning error --- from
\cite{torres-sospedra14:_ujiin} which are based on the distance-based
$k$-Nearest Neighbors ($k$NN) algorithm \cite{cover67:_neares}. As discussed in
Sec.~\ref{sec:scal-dnn-arch}, even though a direct comparison with the results
from the EvAAL/IPIN 2015 competition is not possible due to the lack of testing
samples in the public version of the UJIIndoorLoc dataset and a slightly
different way of calculating the positioning error, our results are also
comparable to the competition results summarized in
Table~\ref{tab:evaal_ipin2015}.

Note that the results presented in this section are not optimized with DNN
parameters, including the number of hidden layers and the number of nodes at
each layer; we investigated the feasibility of the combined
building/floor/location estimation using a single DNN based on multi-label
classification framework with focus on the effects of the number of largest
elements from the output location vector ($\kappa$) and the scaling factor for a
threshold ($\sigma$) in the location coordinates estimation. This leaves much
room for further optimization of the performance.


\section{Conclusions}
\label{sec:conclusions}
In this paper we have proposed a new scalable DNN architecture for
multi-building and multi-floor indoor localization based on Wi-Fi
fingerprinting, which can cover a large-scale complex of many multi-story
buildings under the same management. The proposed DNN architecture consists of
an SAE for the reduction of feature space dimension and a feed-forward
classifier for multi-label classification of
building/floor/location. Reformulating the problem of building/floor/location
classification based on the framework of multi-label classification, we can
achieve better scalability (i.e., greatly reducing the number of DNN output
nodes) and better exploit the hierarchical nature of the building/floor
estimation and the floor-level location coordinates estimation through
systematic label formation (i.e., providing straightforward mapping between a
DNN categorical variable and building, floor \& location vectors) compared to
existing DNN architectures based on multi-class classification.

The experimental results using the UJIIndoorLoc dataset clearly demonstrate the
feasibility of the proposed DNN-based multi-building and multi-floor indoor
localization system, which can provide near state-of-the-art performance using a
single DNN in an integrated way. Combined with the unique advantage of a
DNN-based indoor localization system that, once trained, it does not need the
fingerprint database any longer but carries the necessary information for
localization in DNN weights, the scalable DNN architecture proposed in this
paper could open a door for a future secure and energy-efficient indoor
localization solution exclusively running on mobile devices without exchanging
any data with the server.

Still, there are several areas for further research in multi-building and
multi-floor indoor localization based on DNNs and Wi-Fi fingerprinting. First,
the experimental results described in Sec.~\ref{sec:experimental-results} are
preliminary and not based on optimized DNN parameters; there are still much room
for further optimization of the performance through DNN parameter tuning and the
investigation of the performance and complexity tradeoff. Second, there is no
direct match between the cost function used in DNN training/validation and the
actual performance in the final evaluation for building/floor detection and the
location coordinates estimation due to the additional processing of DNN output
(i.e., the use of $\argmax$ function for the former and the rather complicated
processing with multiple elements of the location vector for the latter). To
fully take into account the actual performance of building/floor detection and
location coordinates estimation during training/validation, we may consider
heuristics like evolutionary algorithms (e.g., genetic algorithm (GA)
\cite{Goldberg:89} and particle swarm optimization (PSO)
\cite{kennedy95:_partic}), simulated annealing \cite{kirkpatrick83:_optim}, and
quantum annealing \cite{finnila94:_quant} for training DNN weights; due to its
many tradeoffs between complexity and flexibility resulting from the use of
heuristics in DNN weight training, this approach could be an interesting topic
for long-term research.


\begin{backmatter}

\section*{Competing interests}
The authors declare that they have no competing interests.

\section*{Author's contributions}
Kyeong Soo Kim, Sanghyuk Lee, and Kaizhu Huang initiated a seed project on the
subject of scalable indoor localization based on DNNs and Wi-Fi fingerprinting,
which this manuscript is based on. Kyeong Soo Kim proposed DNN architectures,
carried out numerical experiments, and drafted the manuscript, which Sanghyuk
Lee and Kaizhu Huang checked and clarified. All authors read and approved the
final manuscript.

\section*{Acknowledgements}
This work was supported in part by Xi'an Jiaotong-Liverpool University (XJTLU)
Research Development Fund (under Grant RDF-14-01-25), Summer Undergraduate
Research Fellowships programme (under Grant SURF-201739), Research Institute for
Smart and Green Cities Seed Grant Programme 2016-2017 (under Grant
RISGC-2017-4), Centre for Smart Grid and Information Convergence, National
Natural Science Foundation of China (NSFC) (under Grant 61473236), Natural
Science Fund for Colleges and Universities in Jiangsu Province (under Grant
17KJD520010), and Suzhou Science and Technology Program (under Grant SYG201712,
SZS201613).




\newcommand{\BMCxmlcomment}[1]{}

\BMCxmlcomment{

<refgrp>

<bibl id="B1">
  <title><p>{Wi-Fi} fingerprint-based indoor positioning: Recent advances and
  comparisons</p></title>
  <aug>
    <au><snm>He</snm><fnm>S</fnm></au>
    <au><snm>Chan</snm><fnm>SHG</fnm></au>
  </aug>
  <source>{IEEE} Commun. Surveys Tuts.</source>
  <pubdate>2016</pubdate>
  <volume>18</volume>
  <issue>1</issue>
  <fpage>466</fpage>
  <lpage>-490</lpage>
</bibl>

<bibl id="B2">
  <title><p>{RADAR}: An in-building {RF}-based user location and tracking
  system</p></title>
  <aug>
    <au><snm>Bahl</snm><fnm>P</fnm></au>
    <au><snm>Padmanabhan</snm><fnm>VN</fnm></au>
  </aug>
  <source>Proc. 2000 IEEE INFOCOM</source>
  <pubdate>2000</pubdate>
  <volume>2</volume>
  <fpage>775</fpage>
  <lpage>-784</lpage>
</bibl>

<bibl id="B3">
  <title><p>{Wi-Fi} fingerprinting in the real world -- {RTLS\@UM} at the
  {EvAAL} competition</p></title>
  <aug>
    <au><snm>Moreira</snm><fnm>A</fnm></au>
    <au><snm>Nicolau</snm><fnm>MJ</fnm></au>
    <au><snm>Meneses</snm><fnm>F</fnm></au>
    <au><snm>Costa</snm><fnm>A</fnm></au>
  </aug>
  <source>Proc. International Conference on Indoor Positioning and Indoor
  Navigation ({IPIN})</source>
  <publisher>Banff, Alberta, Canada</publisher>
  <pubdate>2015</pubdate>
  <fpage>1</fpage>
  <lpage>-10</lpage>
</bibl>

<bibl id="B4">
  <title><p>{UJIIndoorLoc}: A new multi-building and multi-floor database for
  wlan fingerprint-based indoor localization problems</p></title>
  <aug>
    <au><snm>Torres Sospedra</snm><fnm>J</fnm></au>
    <au><snm>Montoliu</snm><fnm>R</fnm></au>
    <au><snm>Mart\'inez Us\'o</snm><fnm>A</fnm></au>
    <au><snm>Avariento</snm><fnm>JP</fnm></au>
    <au><snm>Arnau</snm><fnm>TJ</fnm></au>
    <au><snm>Benedito Bordonau</snm><fnm>M</fnm></au>
    <au><snm>Huerta</snm><fnm>J</fnm></au>
  </aug>
  <source>Proc. International Conference on Indoor Positioning and Indoor
  Navigation ({IPIN})</source>
  <publisher>Busan, Korea</publisher>
  <pubdate>2014</pubdate>
  <fpage>261</fpage>
  <lpage>-270</lpage>
</bibl>

<bibl id="B5">
  <title><p>Deep neural networks for wireless localization in indoor and
  outdoor environments</p></title>
  <aug>
    <au><snm>Zhang</snm><fnm>W</fnm></au>
    <au><snm>Liu</snm><fnm>K</fnm></au>
    <au><snm>Zhang</snm><fnm>W</fnm></au>
    <au><snm>Zhang</snm><fnm>Y</fnm></au>
    <au><snm>Gu</snm><fnm>J</fnm></au>
  </aug>
  <source>Neurocomputing</source>
  <pubdate>2016</pubdate>
  <volume>194</volume>
  <fpage>279</fpage>
  <lpage>-287</lpage>
</bibl>

<bibl id="B6">
  <title><p>A fingerprinting indoor localization algorithm based deep
  learning</p></title>
  <aug>
    <au><snm>F\'elix</snm><fnm>G</fnm></au>
    <au><snm>Siller</snm><fnm>M</fnm></au>
    <au><snm>\'Alvarez</snm><fnm>EN</fnm></au>
  </aug>
  <source>Proc. {ICUFN} 2016</source>
  <publisher>Vienna, Austria</publisher>
  <pubdate>2016</pubdate>
  <fpage>1006</fpage>
  <lpage>-1011</lpage>
</bibl>

<bibl id="B7">
  <title><p>Low-effort place recognition with WiFi fingerprints using deep
  learning</p></title>
  <aug>
    <au><snm>{Nowicki}</snm><fnm>M.</fnm></au>
    <au><snm>{Wietrzykowski}</snm><fnm>J.</fnm></au>
  </aug>
  <source>ArXiv e-prints</source>
  <pubdate>2017</pubdate>
  <url>https://arxiv.org/abs/1611.02049v2</url>
  <note>arXiv:1611.02049v2 [cs.RO]</note>
</bibl>

<bibl id="B8">
  <title><p>Multi-label classification: An overview</p></title>
  <aug>
    <au><snm>Tsoumakas</snm><fnm>G</fnm></au>
    <au><snm>Katakis</snm><fnm>I</fnm></au>
  </aug>
  <source>International Journal of Data Warehousing \& Mining</source>
  <pubdate>2007</pubdate>
  <volume>3</volume>
  <issue>3</issue>
  <fpage>1</fpage>
  <lpage>-13</lpage>
</bibl>

<bibl id="B9">
  <title><p>Multilabel classification</p></title>
  <aug>
    <au><snm>Herrera</snm><fnm>F</fnm></au>
    <au><snm>Charte</snm><fnm>F</fnm></au>
    <au><snm>Rivera</snm><fnm>AJ</fnm></au>
    <au><snm>Jesus</snm><fnm>MJ</fnm></au>
  </aug>
  <source>Multilabel Classification: Problem analysis, metrics and
  techniques</source>
  <publisher>Switzerland: Springer</publisher>
  <section><title><p>2</p></title></section>
  <pubdate>2016</pubdate>
  <fpage>17</fpage>
  <lpage>-31</lpage>
</bibl>

<bibl id="B10">
  <title><p>Smart mobility in smart city: Action taxonomy, {ICT} intensity and
  public benefits</p></title>
  <aug>
    <au><snm>Benevolo</snm><fnm>C</fnm></au>
    <au><snm>Dameri</snm><fnm>RP</fnm></au>
    <au><snm>D'Auria</snm><fnm>B</fnm></au>
  </aug>
  <source>Empowering organizations: Lecture notes in information systems and
  organisation</source>
  <publisher>Switzerland: Springer</publisher>
  <editor>Teresina Torre and Alessio Maria Braccini and Riccardo
  Spinelli</editor>
  <edition>1</edition>
  <series><title><p>Lecture notes in information systems and
  organisation</p></title></series>
  <section><title><p>2</p></title></section>
  <pubdate>2016</pubdate>
  <issue>11</issue>
  <fpage>13</fpage>
  <lpage>-28</lpage>
</bibl>

<bibl id="B11">
  <title><p>Keras: The {P}ython deep learning library</p></title>
  <url>https://keras.io/</url>
</bibl>

<bibl id="B12">
  <title><p>{TensorFlow\textsuperscript{TM}}</p></title>
  <url>https://www.tensorflow.org/</url>
</bibl>

<bibl id="B13">
  <title><p>Nearest neighbor pattern classification</p></title>
  <aug>
    <au><snm>Cover</snm><fnm>T. M.</fnm></au>
    <au><snm>Hart</snm><fnm>P. E.</fnm></au>
  </aug>
  <source>{IEEE} Trans. Inf. Theory</source>
  <pubdate>1967</pubdate>
  <volume>13</volume>
  <issue>1</issue>
  <fpage>21</fpage>
  <lpage>-27</lpage>
</bibl>

<bibl id="B14">
  <title><p>Genetic Algorithms in Search, Optimization and Machine
  Learning</p></title>
  <aug>
    <au><snm>Goldberg</snm><fnm>D. E.</fnm></au>
  </aug>
  <publisher>Reading, Massachusetts: Addison-Wesley</publisher>
  <pubdate>1989</pubdate>
</bibl>

<bibl id="B15">
  <title><p>Particle swarm optimization</p></title>
  <aug>
    <au><snm>Kennedy</snm><fnm>J</fnm></au>
    <au><snm>Eberhart</snm><fnm>R</fnm></au>
  </aug>
  <source>Proc. {IEEE} International Conference on Neural Networks.
  IV.</source>
  <publisher>Perth, WA, Australia</publisher>
  <pubdate>1995</pubdate>
  <fpage>1942</fpage>
  <lpage>-1948</lpage>
</bibl>

<bibl id="B16">
  <title><p>Optimization by simulated annealing</p></title>
  <aug>
    <au><snm>Kirkpatrick</snm><fnm>S.</fnm></au>
    <au><snm>Jr.</snm><fnm>CDG</fnm></au>
    <au><snm>Vecchi</snm><fnm>M. P.</fnm></au>
  </aug>
  <source>Science</source>
  <pubdate>1983</pubdate>
  <volume>220</volume>
  <issue>4598</issue>
  <fpage>671</fpage>
  <lpage>-680</lpage>
</bibl>

<bibl id="B17">
  <title><p>Quantum annealing: A new method for minimizing multidimensional
  functions</p></title>
  <aug>
    <au><snm>Finnila</snm><fnm>A. B.</fnm></au>
    <au><snm>Gomez</snm><fnm>M. A.</fnm></au>
    <au><snm>Sebenik</snm><fnm>C.</fnm></au>
    <au><snm>Stenson</snm><fnm>C.</fnm></au>
    <au><snm>Doll</snm><fnm>J. D.</fnm></au>
  </aug>
  <source>Chemical Physics Letters</source>
  <pubdate>1994</pubdate>
  <volume>219</volume>
  <issue>5-6</issue>
  <fpage>343</fpage>
  <lpage>-348</lpage>
</bibl>

<bibl id="B18">
  <title><p>{ADAM}: A method for stochastic optimization</p></title>
  <aug>
    <au><snm>Kingma</snm><fnm>D</fnm></au>
    <au><snm>Ba</snm><fnm>J</fnm></au>
  </aug>
  <source>ArXiv e-prints</source>
  <pubdate>2017</pubdate>
  <url>https://arxiv.org/abs/1412.6980v8</url>
  <note>arXiv:1412.6980v8 [cs.LG]</note>
</bibl>

</refgrp>
} 



\section*{Figures}

\begin{figure}[!htbp]
  \begin{center}
    \includegraphics[width=.5\linewidth]{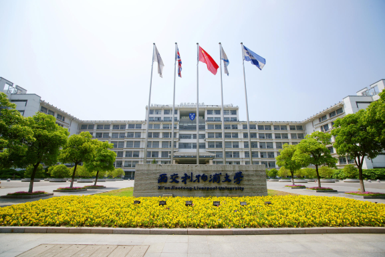}\\
    {\scriptsize (a)}\\
    \includegraphics[width=.9\linewidth]{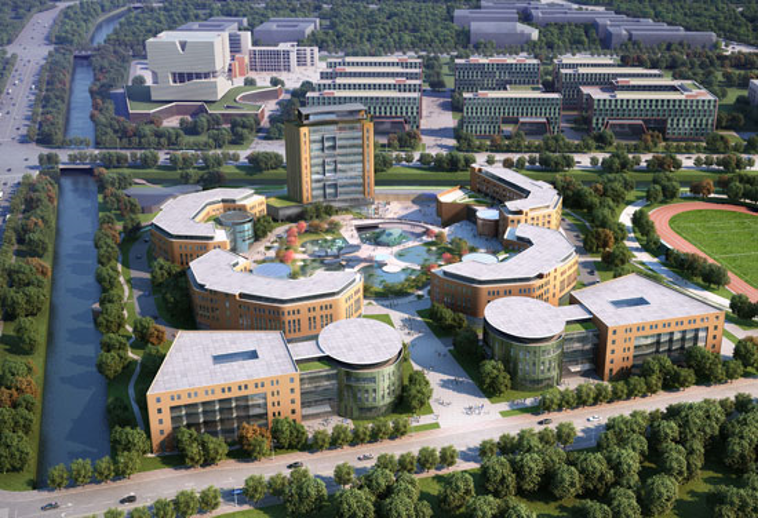}\\
    {\scriptsize (b)}
  \end{center}
  \caption{ XJTLU campus in (a) 2006 and (b) 2017.}
  \label{fig:xjtlu_campus}
\end{figure}

\begin{figure}[!htbp]
  \begin{center}
    \includegraphics[angle=-90,width=.75\linewidth]{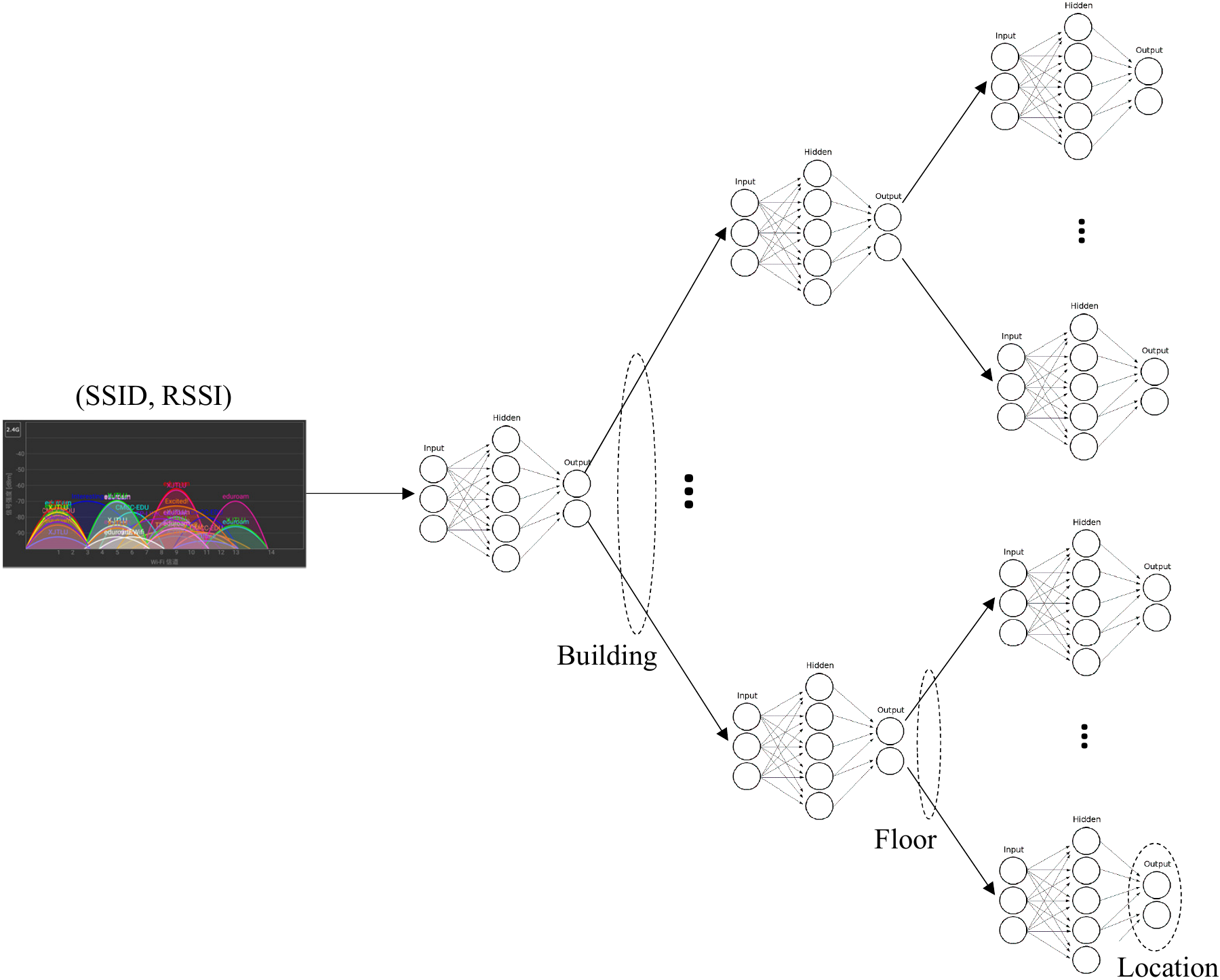}\\
    {\scriptsize (a)}\\
    \includegraphics[angle=-90,width=.95\linewidth]{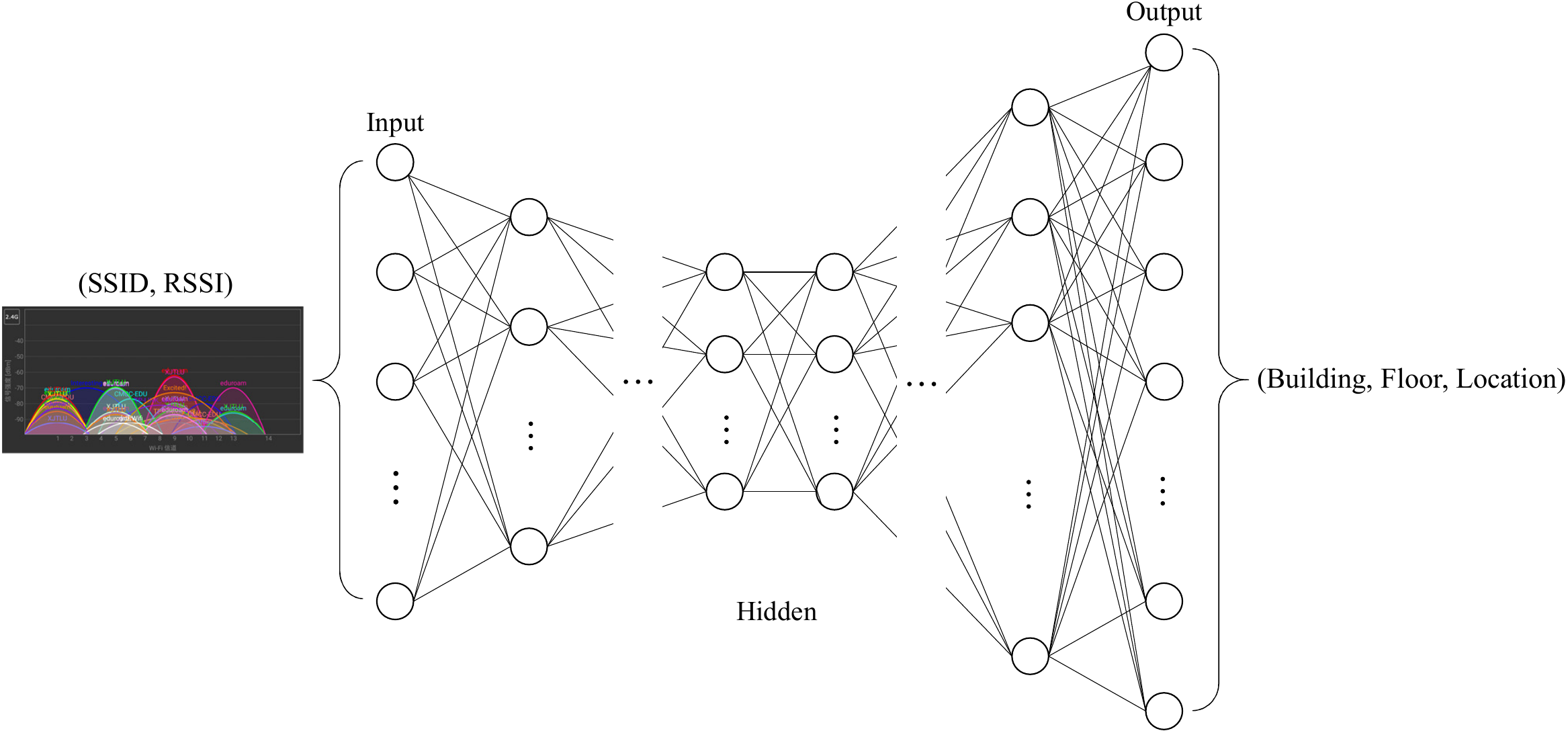}\\
    {\scriptsize (b)}
  \end{center}
  \caption{System architectures for large-scale DNN-based multi-building and
    multi-floor indoor localization: (a) A hierarchical architecture; (b) an
    integrated architecture.}
  \label{fig:hierarchical_vs_integrated}
\end{figure}

\begin{figure}[!htbp]
  \begin{center}
    \includegraphics[angle=-90,width=.7\linewidth]{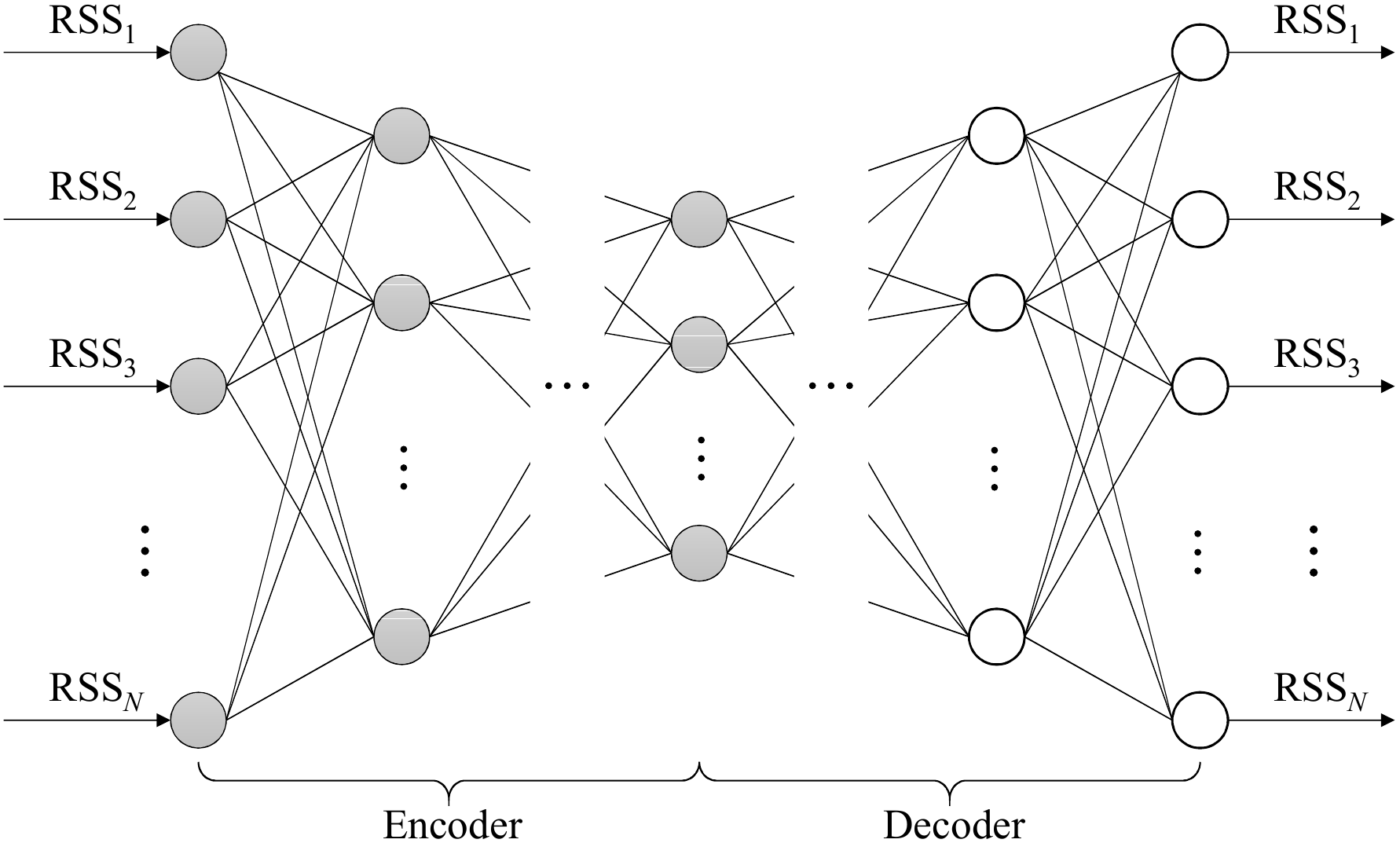}\\
    {\scriptsize (a)}\\
    \includegraphics[angle=-90,width=.9\linewidth]{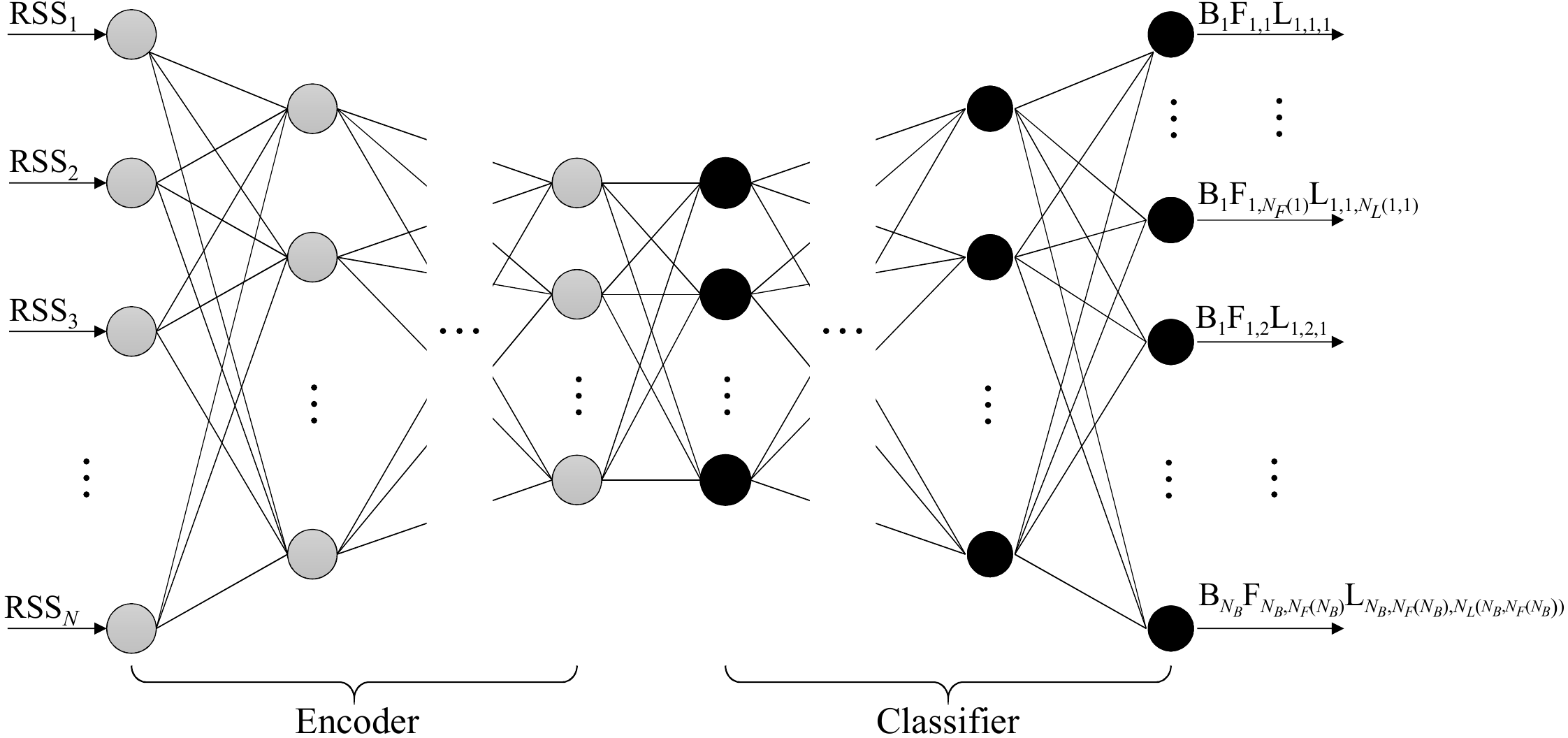}\\
    {\scriptsize (b)}
  \end{center}
  \caption{A DNN architectures for the combined estimation of building, floor,
    and location: (a) A stacked autoencoder (SAE) for the reduction of feature
    space dimension; (b) an overall architecture consisting of the encoder part
    of the SAE and a feed-forward classifier for multi-class classification with
    flattened building-floor-location labels as in \cite{nowicki17:_low_wifi}.}
  \label{fig:flat_dnn_classifier}
\end{figure}

\begin{figure}[!htbp]
  \begin{center}
    \includegraphics[angle=-90,width=.9\linewidth]{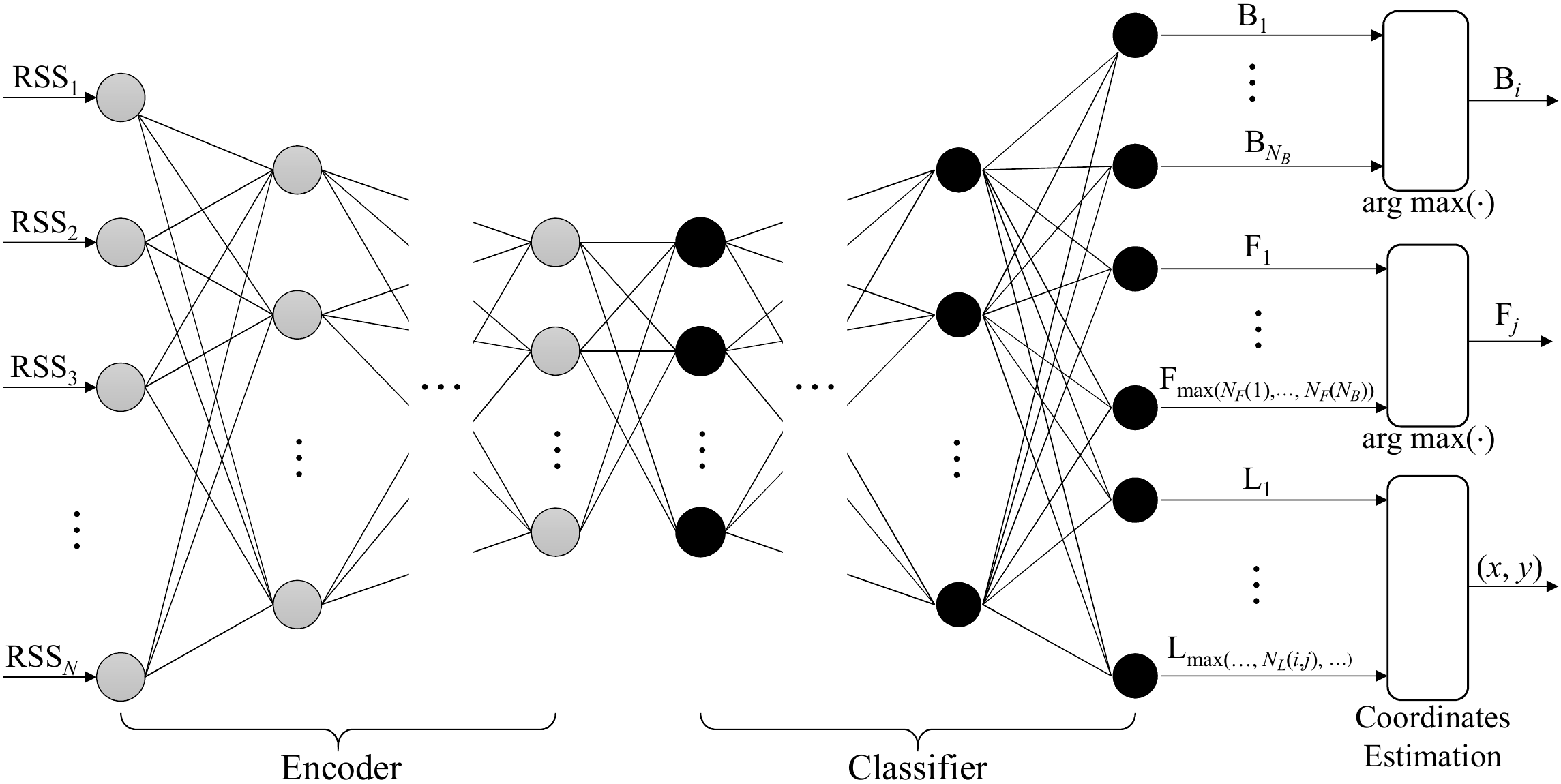}
  \end{center}
  \caption{A DNN architectures for scalable building/floor classification and
    floor-level coordinates estimation based on an SAE for the reduction of
    feature space dimension and a feed-forward classifier for multi-label
    classification.}
  \label{fig:scalable_dnn_classifier}
\end{figure}

\begin{algorithm}[!htbp]
  \DontPrintSemicolon \SetKwInOut{Input}{Input}%
  \SetKwInOut{Output}{Output}%
  \Input{Training/validation dataset $\boldsymbol{\mathcal{D}}$}%
  \Input{Estimated building ID B$_{i}$}%
  \Input{Estimated floor ID F$_{j}$}%
  \Input{$\boldsymbol{\mathcal{L}}=\left\{\mathcal{L}_{1} \ldots
      \mathcal{L}_{\kappa}\right\}$, a set of $\kappa$ largest elements of the
    location vector $\boldsymbol{L}$ from the DNN classifier}%
  \Output{Estimated location coordinates $\boldsymbol{C_{s}}=(x_{s}, y_{s})$
    (\textit{centroid})}%
  \Output{Estimated location coordinates $\boldsymbol{C_{w}}=(x_{w}, y_{w})$
    (\textit{weighted centroid})}%
  \BlankLine%
  $sum_{x} \leftarrow 0$\;%
  $wsum_{x} \leftarrow 0$\;%
  $sum_{y} \leftarrow 0$\;%
  $wsum_{y} \leftarrow 0$\;%
  $sum_{w} \leftarrow 0$\;%
  $n_{c} \leftarrow 0$ \tcp*[h]{number of candidate coordinates}\;
  \For{$l \in \boldsymbol{\mathcal{L}}$}{%
    \If(\tcp*[h]{threshold
      checking}){$l \geq \sigma{\times}\max(\boldsymbol{L})$}{%
      \If(\tcp*[h]{i.e., a valid location of the
        dataset}){$(\text{B}_{i}, \text{F}_{j}, l) \in
        \boldsymbol{\mathcal{D}}$}{%
        $x \leftarrow \coords_{x}(\text{B}_{i}, \text{F}_{j}, l)$
        \tcp*[h]{return the $x$ coordinate of building/floor/location}\;
        $y \leftarrow \coords_{y}(\text{B}_{i}, \text{F}_{j}, l)$
        \tcp*[h]{return the $y$ coordinate of building/floor/location}\;
        $sum_{x} \leftarrow sum_{x} + x$\;
        $wsum_{x} \leftarrow wsum_{x} + l {\times} x$\;
        $sum_{y} \leftarrow sum_{y} + y$\;
        $wsum_{y} \leftarrow wsum_{y} + l {\times} y$\;
        $sum_{w} \leftarrow sum_{w} + l$\; $n_{c} \leftarrow n_{c} + 1$\;%
      }%
    }%
  }%
  \If{$n_{c} > 0$}{%
    $\boldsymbol{C_{s}} \leftarrow (\frac{sum_{x}}{n_{c}},
    \frac{sum_{y}}{n_{c}})$\;
    $\boldsymbol{C_{w}} \leftarrow (\frac{wsum_{x}}{sum_{w}},
    \frac{wsum_{y}}{sum_{w}})$\;%
  }%
  \Else(\tcp*[h]{i.e., no candidate coordinates}){%
    $\boldsymbol{C_{s}}, \boldsymbol{C_{w}} \leftarrow \text{centroid of
      \textit{all} coordinates belonging to B$_{i}$ and F$_{j}$ of
      $\boldsymbol{\mathcal{D}}$}$%
  }%
  \caption{Procedure for the estimation of location coordinates.}
  \label{fig:coordinates_estimation}
\end{algorithm}

\begin{figure}[!htbp]
  \begin{center}
    \includegraphics[angle=-90,width=.7\linewidth]{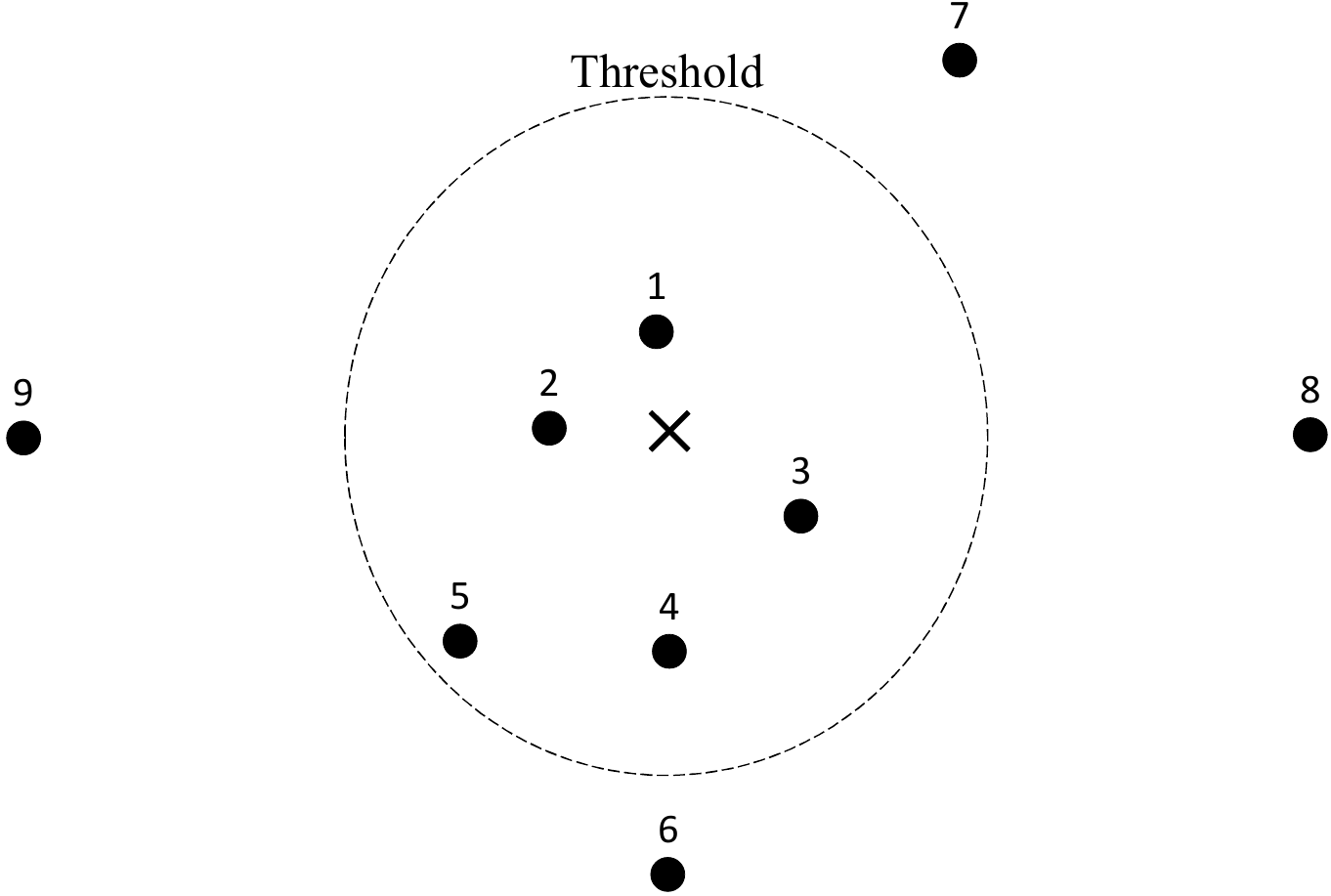}\\
    \vspace{0.2cm}
    {\scriptsize (a)}\\
    \includegraphics[angle=-90,width=.7\linewidth]{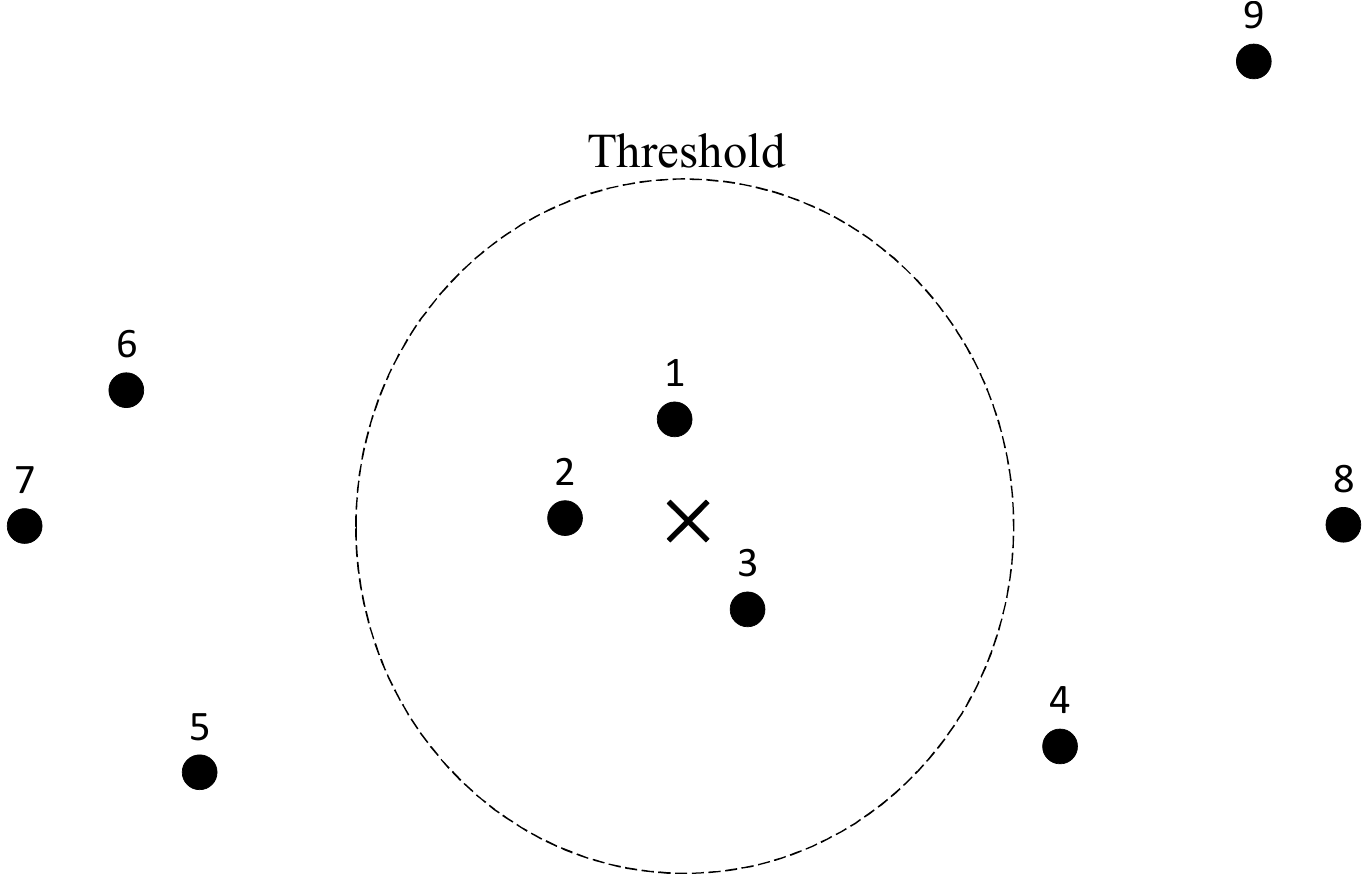}\\
    \vspace{0.2cm}
    {\scriptsize (b)}
  \end{center}
  \caption{Two examples of location coordinates estimation, where
    $\boldsymbol{\times}$ denotes a new location whose coordinates are to be
    estimated and $\CIRCLE$ denotes a reference point whose coordinates are
    stored in the fingerprint database. The numbers over the reference points
    indicate their closeness to the new location with 1 being the closest.}
  \label{fig:coordinates_examples}
\end{figure}


\section*{Tables}

\begin{table*}[!htbp]
  \caption{Label formation example for the multi-label classification of
    building, floor, and location with two multi-story buildings.}
  \label{tab:label_formation}
  \centering
  \begin{tabular}{ccccc}
    \hline
    Building & Floor & Location & Sequential Coding & One-Hot Coding \\
    \hline
    \multirow{7}{*}{A} & \multirow{2}{*}{1st} & Lecture Theater A & $0,0,0$ & $01|0001|0001$
    \\
             & & Lecture Theater B & $0,0,1$ & $01|0001|0010$ \\ \cline{2-5}
             & \multirow{2}{*}{2nd} & Lab 1 & $0,1,0$ & $01|0010|0001$ \\
             & & Lab 2 & $0,1,1$ & $01|0010|0010$ \\ \cline{2-5}
             & \multirow{3}{*}{3rd} & A301 & $0,2,0$ & $01|0100|0001$ \\
             & & A302 & $0,2,1$ & $01|0100|0010$ \\
             & & A303 & $0,2,2$ & $01|0100|0100$ \\ \hline
    \multirow{9}{*}{B} & \multirow{2}{*}{1st} & Common Room & $1,0,0$ & $10|0001|0001$
    \\
             & & Printing Room & $1,0,1$ & $10|0001|0010$ \\ \cline{2-5}
             & \multirow{2}{*}{2nd} & Conference Room 1 & $1,1,0$ & $10|0010|0001$ \\
             & & Conference Room 2 & $1,1,1$ & $10|0010|0010$ \\ \cline{2-5}
             & \multirow{4}{*}{3rd} & B301 & $1,2,0$ & $10|0100|0001$ \\
             & & B302 & $1,2,1$ & $10|0100|0010$ \\
             & & B303 & $1,2,2$ & $10|0100|0100$ \\
             & & B304 & $1,2,3$ & $10|0100|1000$ \\ \cline{2-5}
             & 4th & Gym & $1,3,0$ & $10|1000|0001$ \\ \hline
  \end{tabular}
\end{table*}

\begin{table}[!htbp]
  \caption{Parameter Values for Scalable DNN-Based Indoor Localization.}
  \label{tab:scalable_localization_dnn_parameters}
  \centering
  \begin{tabular}{cc}
    \hline
    DNN Parameter & Value \\ \hline
    Ratio of Training Data to Overall Data & 0.90 \\
    Number of Epochs & 20 \\
    Batch Size & 10 \\ \Xhline{0.4\arrayrulewidth}
    SAE Hidden Layers & 256-128-256 \\
    SAE Activation & Rectified Linear (ReLU) \\
    SAE Optimizer & ADAM \cite{kingma17:_adam} \\
    SAE Loss & Mean Squared Error (MSE) \\ \Xhline{0.4\arrayrulewidth}
    Classifier Hidden Layers & 64-128 \\
    Classifier Activation & ReLU \\
    Classifier Optimizer & ADAM \\
    Classifier Loss & Binary Crossentropy \\
    Classifier Dropout Rate & 0.20 \\ \hline
  \end{tabular}
\end{table}

\begin{table*}[!htbp]
  \begin{threeparttable}
    \sisetup{               
      detect-family,
      round-mode=places,
      round-precision=2,
      scientific-notation = fixed,
      fixed-exponent = 0
    }%
    \definecolor{light-gray}{gray}{0.75}
    \caption{Effects of the number of largest elements from the output location
      vector ($\kappa$) and the scaling factor for a threshold ($\sigma$) on the
      Performance of multi-building and multi-floor indoor localization.}
    \label{tab:scalable_localization_performance}
    \centering
    \begin{tabular}{ccccccc}
      \hline
      \multirow{2}{*}{$\kappa$} & \multirow{2}{*}{$\sigma$} & \multirow{2}{*}{Building Hit Rate [\%]} & \multirow{2}{*}{Floor Hit Rate [\%]} & \multirow{2}{*}{Success Rate [\%]} & \multicolumn{2}{c}{Positioning Error [m]} \\ \cline{6-7}
                                & & & & & Centroid & Weighted Centroid \\ \hline
      1 & N/A\tnote{*} & \num{9.981998e+01} & \num{9.189919e+01} & \num{9.180918e+01} & \num{1.140346e+01} & \num{1.140346e+01} \\ \hline
      \multirow{6}{*}{2} & 0.0 & \num{9.936994e+01} & \num{9.243924e+01} & \num{9.180918e+01} & \num{1.062041e+01} & \num{1.053982e+01} \\
                                & 0.1 & \num{1.000000e+02} & \num{9.180918e+01} & \num{9.180918e+01} & \num{1.040084e+01} & \num{1.032711e+01} \\
                                & 0.2 & \cellcolor{light-gray} \num{9.981998e+01} & \cellcolor{light-gray} \num{9.261926e+01} & \cellcolor{light-gray} \num{9.243924e+01} & \cellcolor{light-gray} \num{9.739654e+00} & \cellcolor{light-gray} \num{9.661333e+00} \\
                                & 0.3 & \cellcolor{light-gray} \num{9.963996e+01} & \cellcolor{light-gray} \num{9.198920e+01} & \cellcolor{light-gray} \num{9.180918e+01} & \cellcolor{light-gray} \num{9.777987e+00} & \cellcolor{light-gray} \num{9.710835e+00} \\
                                & 0.4 & \num{9.972997e+01} & \num{9.153915e+01} & \num{9.144914e+01} & \num{1.028777e+01} & \num{1.020546e+01} \\
                                & 0.5 & \num{1.000000e+02} & \num{9.000900e+01} & \num{9.000900e+01} & \num{1.015949e+01} & \num{1.009243e+01} \\ \hline
      \multirow{6}{*}{3} & 0.0 & \num{9.972997e+01} & \num{9.153915e+01} & \num{9.135914e+01} & \num{1.014222e+01} & \num{9.792770e+00} \\
                                & 0.1 & \cellcolor{light-gray} \num{9.990999e+01} & \cellcolor{light-gray} \num{9.090909e+01} & \cellcolor{light-gray} \num{9.081908e+01} & \cellcolor{light-gray} \num{9.918299e+00} & \cellcolor{light-gray} \num{9.758788e+00} \\
                                & 0.2 & \cellcolor{light-gray} \num{9.882988e+01} & \cellcolor{light-gray} \num{9.090909e+01} & \cellcolor{light-gray} \num{9.027903e+01} & \cellcolor{light-gray} \num{9.983135e+00} & \cellcolor{light-gray} \num{9.795777e+00} \\
                                & 0.3 & \num{9.954995e+01} & \num{9.207921e+01} & \num{9.189919e+01} & \num{1.013391e+01} & \num{1.000786e+01} \\
                                & 0.4 & \num{9.990999e+01} & \num{9.198920e+01} & \num{9.198920e+01} & \num{1.063061e+01} & \num{1.047131e+01} \\
                                & 0.5 & \cellcolor{light-gray} \num{9.981998e+01} & \cellcolor{light-gray} \num{9.036904e+01} & \cellcolor{light-gray} \num{9.036904e+01} & \cellcolor{light-gray} \num{9.937904e+00} & \cellcolor{light-gray} \num{9.889872e+00} \\ \hline
    \multirow{6}{*}{4} & 0.0 & \num{9.981998e+01} & \num{9.090909e+01} & \num{9.090909e+01} & \num{1.027489e+01} & \num{9.662943e+00} \\
                                & 0.1 & \num{9.936994e+01} & \num{9.198920e+01} & \num{9.162916e+01} & \num{1.036905e+01} & \num{9.920916e+00} \\
                                & 0.2 & \num{9.963996e+01} & \num{9.207921e+01} & \num{9.189919e+01} & \num{1.025626e+01} & \num{1.008654e+01} \\
                                & 0.3 & \num{9.981998e+01} & \num{9.144914e+01} & \num{9.135914e+01} & \num{1.023558e+01} & \num{1.015537e+01} \\
                                & 0.4 & \num{9.990999e+01} & \num{9.225923e+01} & \num{9.216922e+01} & \num{1.034791e+01} & \num{1.023167e+01} \\
                                & 0.5 & \num{9.981998e+01} & \num{9.126913e+01} & \num{9.117912e+01} & \num{1.009757e+01} & \num{1.007306e+01} \\ \hline
      \multirow{6}{*}{5} & 0.0 & \num{9.990999e+01} & \num{9.135914e+01} & \num{9.126913e+01} & \num{1.128615e+01} & \num{1.036350e+01} \\
                                & 0.1 & \cellcolor{light-gray} \num{9.990999e+01} & \cellcolor{light-gray} \num{9.162916e+01} & \cellcolor{light-gray} \num{9.162916e+01} & \cellcolor{light-gray} \num{9.897813e+00} & \cellcolor{light-gray} \num{9.618697e+00} \\
                                & 0.2 & \cellcolor{light-gray} \num{9.990999e+01} & \cellcolor{light-gray} \num{9.072907e+01} & \cellcolor{light-gray} \num{9.072907e+01} & \cellcolor{light-gray} \num{9.887615e+00} & \cellcolor{light-gray} \num{9.571252e+00} \\
                                & 0.3 & \num{9.981998e+01} & \num{9.090909e+01} & \num{9.081908e+01} & \num{1.026534e+01} & \num{9.992006e+00} \\
                                & 0.4 & \num{9.972997e+01} & \num{9.216922e+01} & \num{9.207921e+01} & \num{1.016706e+01} & \num{1.001166e+01} \\
                                & 0.5 & \num{9.981998e+01} & \num{9.297930e+01} & \num{9.288929e+01} & \num{1.059074e+01} & \num{1.053813e+01} \\ \hline
      \multirow{6}{*}{6} & 0.0 & \num{9.981998e+01} & \num{9.189919e+01} & \num{9.171917e+01} & \num{1.084073e+01} & \num{9.713978e+00} \\
                                & 0.1 & \num{9.963996e+01} & \num{9.207921e+01} & \num{9.180918e+01} & \num{1.035306e+01} & \num{9.861424e+00} \\
                                & 0.2 & \cellcolor{light-gray} \num{1.000000e+02} & \cellcolor{light-gray} \num{9.198920e+01} & \cellcolor{light-gray} \num{9.198920e+01} & \cellcolor{light-gray} \num{9.853155e+00} & \cellcolor{light-gray} \num{9.564450e+00} \\
                                & 0.3 & \num{9.981998e+01} & \num{9.279928e+01} & \num{9.279928e+01} & \num{1.048978e+01} & \num{1.022295e+01} \\
                                & 0.4 & \num{9.936994e+01} & \num{9.108911e+01} & \num{9.099910e+01} & \num{1.032216e+01} & \num{1.017355e+01} \\
                                & 0.5 & \cellcolor{light-gray} \num{9.963996e+01} & \cellcolor{light-gray} \num{9.090909e+01} & \cellcolor{light-gray} \num{9.063906e+01} & \cellcolor{light-gray} \num{9.554220e+00} & \cellcolor{light-gray} \num{9.519261e+00} \\ \hline
      \multirow{6}{*}{7} & 0.0 & \num{9.981998e+01} & \num{8.928893e+01} & \num{8.928893e+01} & \num{1.174417e+01} & \num{1.022120e+01} \\
                                & 0.1 & \num{9.981998e+01} & \num{9.018902e+01} & \num{9.000900e+01} & \num{1.042525e+01} & \num{9.820845e+00} \\
                                & 0.2 & \num{9.990999e+01} & \num{9.144914e+01} & \num{9.144914e+01} & \num{1.000036e+01} & \num{9.551645e+00} \\
                                & 0.3 & \cellcolor{light-gray} \num{9.990999e+01} & \cellcolor{light-gray} \num{9.162916e+01} & \cellcolor{light-gray} \num{9.153915e+01} & \cellcolor{light-gray} \num{9.746898e+00} & \cellcolor{light-gray} \num{9.533490e+00} \\
                                & 0.4 & \num{9.963996e+01} & \num{9.045905e+01} & \num{9.018902e+01} & \num{1.041638e+01} & \num{1.028135e+01} \\
                                & 0.5 & \cellcolor{light-gray} \num{9.954995e+01} & \cellcolor{light-gray} \num{9.144914e+01} & \cellcolor{light-gray} \num{9.135914e+01} & \cellcolor{light-gray} \num{9.830048e+00} & \cellcolor{light-gray} \num{9.733772e+00} \\ \hline
      \multirow{6}{*}{8} & 0.0 & \num{9.990999e+01} & \num{9.018902e+01} & \num{9.009901e+01} & \num{1.131833e+01} & \num{9.270887e+00} \\
                                & 0.1 & \num{1.000000e+02} & \num{9.126913e+01} & \num{9.126913e+01} & \num{1.061573e+01} & \num{1.014184e+01} \\
                                & 0.2 & \cellcolor{gray} \num{9.981998e+01} & \cellcolor{gray} \num{9.126913e+01} & \cellcolor{gray} \num{9.117912e+01} &  \cellcolor{gray} \num{9.755503e+00} & \cellcolor{gray} \num{9.285992e+00} \\
                                & 0.3 & \cellcolor{light-gray} \num{9.981998e+01} & \cellcolor{light-gray} \num{9.054905e+01} & \cellcolor{light-gray} \num{9.036904e+01} & \cellcolor{light-gray} \num{9.953853e+00} & \cellcolor{light-gray} \num{9.819975e+00} \\
                                & 0.4 & \num{9.990999e+01} & \num{9.036904e+01} & \num{9.027903e+01} & \num{1.021126e+01} & \num{1.013918e+01} \\
                                & 0.5 & \cellcolor{light-gray} \num{9.990999e+01} & \cellcolor{light-gray} \num{9.054905e+01} & \cellcolor{light-gray} \num{9.054905e+01} & \cellcolor{light-gray} \num{9.861089e+00} & \cellcolor{light-gray} \num{9.794417e+00} \\ \hline
      \multirow{6}{*}{9} & 0.0 & \num{9.981998e+01} & \num{9.099910e+01} & \num{9.090909e+01} & \num{1.274973e+01} & \num{9.758554e+00} \\
                                & 0.1 & \num{9.981998e+01} & \num{9.027903e+01} & \num{9.018902e+01} & \num{1.016850e+01} & \num{9.548974e+00} \\
                                & 0.2 & \num{9.981998e+01} & \num{9.072907e+01} & \num{9.063906e+01} & \num{1.005135e+01} & \num{9.760677e+00} \\
                                & 0.3 & \num{9.981998e+01} & \num{9.144914e+01} & \num{9.135914e+01} & \num{1.035995e+01} & \num{1.023228e+01} \\
                                & 0.4 & \cellcolor{light-gray} \num{1.000000e+02} & \cellcolor{light-gray} \num{9.180918e+01} & \cellcolor{light-gray} \num{9.180918e+01} & \cellcolor{light-gray} \num{9.862838e+00} & \cellcolor{light-gray} \num{9.784878e+00} \\
                                & 0.5 & \num{9.990999e+01} & \num{9.279928e+01} & \num{9.270927e+01} & \num{1.026204e+01} & \num{1.022622e+01} \\ \hline
      \multirow{6}{*}{10} & 0.0 & \num{9.990999e+01} & \num{9.090909e+01} & \num{9.081908e+01} & \num{1.321298e+01} & \num{9.750404e+00} \\
                                & 0.1 & \num{9.981998e+01} & \num{9.270927e+01} & \num{9.261926e+01} & \num{1.037180e+01} & \num{9.849004e+00} \\
                                & 0.2 & \num{9.990999e+01} & \num{9.171917e+01} & \num{9.171917e+01} & \num{1.033383e+01} & \num{1.003775e+01} \\
                                & 0.3 & \num{1.000000e+02} & \num{9.081908e+01} & \num{9.081908e+01} & \num{1.047165e+01} & \num{1.023868e+01} \\
                                & 0.4 & \num{1.000000e+02} & \num{9.108911e+01} & \num{9.108911e+01} & \num{1.010107e+01} & \num{9.957454e+00} \\
                                & 0.5 & \cellcolor{light-gray} \num{1.000000e+02} & \cellcolor{light-gray} \num{9.189919e+01} & \cellcolor{light-gray} \num{9.189919e+01} & \cellcolor{light-gray} \num{9.955224e+00} & \cellcolor{light-gray} \num{9.913192e+00} \\ \hline
    \end{tabular}
    \begin{tablenotes}
    \item[*] N/A = not applicable; when $\kappa{=}1$, the value of $\sigma$ does
      not affect the selection of locations (i.e., reference points) included in
      the coordinates estimation.
    \end{tablenotes}
  \end{threeparttable}
\end{table*}

\begin{table}[!htpb]
  \caption{Best results from the four teams at the EvAAL/IPIN 2015 competition
    \cite{moreira15:_wi_fi}.}
  \label{tab:evaal_ipin2015}
  \centering
  \begin{tabular}{lcccc}
    \hline
    & MOSAIC & HFTS & RTLS\@UM & ICSL \\ \hline
    Building Hit Rate [\%] & 98.65 & 100 & 100 & 100 \\
    Floor Hit Rate [\%]& 93.86 & 96.25 & 93.74 & 86.93 \\
    Positioning Error (Mean) [m] & 11.64 & 8.49 & 6.20 & 7.67 \\
    Positioning Error (Median) [m] & 6.7 & 7.0 & 4.6 & 5.9 \\ \hline
  \end{tabular}
\end{table}




\end{backmatter}
\end{document}